\documentclass[aip, amsmath,amssymb,preprint]{revtex4-1}

\usepackage{graphicx}
\usepackage{dcolumn}
\usepackage{bm}
\usepackage[english]{babel}
\usepackage[utf8]{inputenc}
\usepackage[T1]{fontenc}
\usepackage{mathptmx}
\usepackage[version=4]{mhchem}
\usepackage[loose]{units}
\usepackage{subfigure}
\begin{document}

\preprint{AIP/123-QED}

\title[Detecting coherent core-hole wave-packet dynamics in \ce{N2}]{Detecting coherent core-hole wave-packet dynamics in \ce{N2} by time- and angle-resolved inner-shell photoelectron spectroscopy}

\author{Ludger~Inhester}
\email{ludger.inhester@cfel.de}
\affiliation{Center for Free-Electron Laser Science, DESY, Notkestrasse 85, 22607 Hamburg, Germany}

\author{Loren~Greenman}
\affiliation{Department of Physics, Kansas State University, Manhattan, Kansas 66506, USA}

\author{Artem~Rudenko}
\affiliation{Department of Physics, Kansas State University, Manhattan, Kansas 66506, USA}

\author{Daniel~Rolles}
\affiliation{Department of Physics, Kansas State University, Manhattan, Kansas 66506, USA}

\author{Robin~Santra}
\affiliation{Center for Free-Electron Laser Science, DESY, Notkestrasse 85, 22607 Hamburg, Germany}
\affiliation{Department of Physics, Universität Hamburg, Jungiusstrasse 9, 20355 Hamburg, Germany}%

\date{\today}

\begin{abstract}
We propose an imaging technique to follow core-hole wave-packet oscillations in the nitrogen molecule.
In this scheme, an attosecond x-ray pulse core-ionizes the nitrogen molecule
and a subsequent attosecond x-ray pulse probes the evolution of the electron dynamics.
We can image the oscillation of the core-hole between the two atomic sites
by measuring the angular correlation between photoelectrons.
Analytical relations for the angular correlation are derived 
based on the plane-wave approximation for the photoelectron wave function.
We validate these results with a scattering calculation for the photoelectron wave function.
The feasibility of the experimental realization of this scheme is discussed
in light of current and future capabilities of x-ray free-electron lasers.
\end{abstract}

\maketitle
\section{Introduction}

The movement of electrons in atoms or molecules is enormously fast and 
typically characterized by an attosecond time scale.
With the technical advances in the field of high-harmonic generation and the creation of attosecond pulses\cite{krausz_attosecond_2009,corkum_attosecond_2007,chang_fundamentals_2016},
the time resolution required to probe coherent dynamics in atoms and molecules has become available
\cite{smirnova_high_2009,sansone_electron_2010,goulielmakis_realtime_2010,haessler_attosecond_2010,tzallas_extremeultraviolet_2011,
calegari_ultrafast_2014,kraus_measurement_2015}.
Besides the fundamental understanding of such dynamics in the context of light-matter interaction, 
studying the coherent electron dynamics in molecules is of particular interest
because it may pave the way for a detailed investigation of the 
time-dependent charge migration in a molecule\cite{kraus_measurement_2015} and may provide means to control chemical dynamics\cite{kling_sub_2013,golubev_control_2015}.
The motion of an electronic wave packet in a molecule has therefore attracted a lot of attention in theoretical works\cite{cederbaum_ultrafast_1999,breidbach_migration_2003,kuleff_multielectron_2005,remacle_electronic_2006,lunnemann_ultrafast_2008,despre_attosecond_2015,kuleff_core_2016,picon_augerinduced_2018}.

Besides the challenge to create extremely short light pulses that are 
able to trigger such ultrafast dynamics, 
the probing of coherent electron dynamics and the corresponding
interpretation of measurement signals is often not trivial.
The question on how to properly image 
the electron dynamics in a molecule or atom 
in a nonstationary electronic state 
has been investigated in several theoretical works,
specifically, in the context of x-ray scattering\cite{dixit_imaging_2012,dixit_theory_2014,popova-gorelova_imaging_2015,popova-gorelova_imaging_2015a,grosser_attosecond_2017,simmermacher_electronic_2019}, 
multi-dimensional spectroscopy\cite{mukamel_multidimensional_2013,biggs_watching_2013,ye_imaging_2019},
transient x-ray absorption spectroscopy\cite{hollstein_correlationdriven_2017}, 
electron diffraction\cite{yuan_exploring_2017},
Auger electron spectroscopy\cite{cooper_singlephoton_2013,cooper_analysis_2014}, and
molecular-frame photoelectron spectroscopy\cite{mignolet_localized_2012,popova-gorelova_imaging_2016,yuan_timeresolved_2018,yuan_ultrafast_2019}.
All these studies have addressed electron wave-packet dynamics in the outer-valence electron shells.
Recent improvements of x-ray free-electron lasers have made it possible to create x-ray pulses with pulse durations below 1 femtosecond\cite{huang_generating_2017}, which can be focused tightly to provide extremely high intensities sufficient for
x-ray multiphoton ionization.
Furthermore, x-ray pulses can be generated with two colors in a two-pulse sequence with adjustable time delay\cite{lutman_experimental_2013,picon_heterositespecific_2016}.
These capabilities open up new routes to trigger and probe coherent electronic wave-packet dynamics in the x-ray regime.
One of the new prospective applications in this context is the probing of electronic wave packet dynamics occurring in 
molecular inner shells and core levels.

In this work, we explore new opportunities provided by x-ray free-electron lasers to measure coherent core-hole
wave-packet dynamics triggered by core ionization of the nitrogen
molecule (\ce{N2}).
We propose to probe these dynamics by subsequently ionizing the \ce{N2+}[$1s^{-1}$] molecular ion into a double-core-ionized state.
As we demonstrate here, by 
detecting both photoelectrons and their angular distribution, 
the motion of the core-hole wave packet can be monitored.

The two core orbitals in \ce{N2}, $1 \sigma_g$ and $1 \sigma_u$, have an energy separation of
$\Delta \epsilon_{gu}=\epsilon_{1\sigma_u}-\epsilon_{1\sigma_g} \simeq \unit[100]{meV}$\cite{hergenhahn_symmetryselective_2001,rolles_isotopeinduced_2005,ehara_symmetrydependent_2006,puttner_statedependent_2008}. 
The delocalized character of the core hole in \ce{N2} has been probed before via molecular-frame angle-resolved 
photoelectron spectroscopy\cite{rolles_isotopeinduced_2005}. 
Moreover, angle-resolved coincident detection of photo- and Auger electrons 
has shown that the core hole in \ce{N2} is in general a superposition of holes in both core orbitals,
and the specific character of the superposition 
depends on the emission direction of the photoelectron\cite{schoffler_ultrafast_2008}.
In general, ionization of the core levels in \ce{N2} by a pulse with coherent bandwidth 
larger than the energy separation $\Delta \epsilon_{gu}$ can lead
to a coherent superposition of two electronic eigenstates that evolves in time.
Accordingly, ionization of \ce{N2} by an x-ray pulse with a duration of a few hundred attoseconds triggers an oscillation in the 
core-ionized molecule with a period of $T=h/\Delta \epsilon_{gu} \simeq \unit[41.4]{fs}$ that periodically
shifts the core vacancy from one atom to the other.
At the same time one must be aware that the core-hole state decays most likely via Auger decay and has a lifetime of $\simeq \unit[7]{fs}$\cite{hergenhahn_symmetryselective_2001}.
Any core-hole wave-packet dynamics in \ce{N2+}[$1s^{-1}$] is therefore strongly suppressed at larger times and only 
the first half period of such dynamics is therefore of relevance.
By core-ionizing the molecule a second time we propose to probe these dynamics and thereby 
obtain a time-dependent image of the core-hole changing its character from being localized on 
one atom to being delocalized on both atoms.

The paper is organized as follows:
In Sec.~\ref{sec:theory}, we introduce the theoretical framework, in which we describe the sequential ionization, and
derive working equations for the angular distribution of the two K-shell photoelectrons based on the plane-wave approximation for the photoelectrons.
In Sec.~\ref{sec:numeric}, we investigate numerical results for suggested pulse parameters and angular directions and discuss the validity of the plane-wave approximation by comparing with a scattering calculation.
In Sec.~\ref{sec:realization},
the feasibility of the proposed experiment is discussed in light of the capabilities of current and future
x-ray free-electron laser facilities, and in Sec.~\ref{sec:conclusion}, we draw final conclusions.

\section{Theory\label{sec:theory}}
Since the interaction of the x-ray light with the molecule 
is considerable weaker than the electronic ionization potential,  
we describe the interaction of the x-ray light with the \ce{N2} molecule
perturbatively.
To that end, we divide the total Hamiltonian $H$ into
\begin{equation}
H=H_0+H_I(t),
\end{equation}
where $H_0$ is the molecular Hamiltonian and
$H_I(t)$ is the interaction between the molecule and the electromagnetic field,
\begin{equation}
H_I(t)=\alpha \int d^3 r \psi^\dagger({\bf r})  \,   \textbf{p} \cdot \textbf{A}({\bf r},t) \, \psi({\bf r}),
\end{equation}
where $\textbf{A}({\bf r},t)$ is the vector potential at position {\bf r} and at time $t$, ${\bf p}$ the canonical electron momentum operator, $\psi^\dagger({\bf r})$ [$\psi({\bf r})$] is a fermionic field operator that creates (annihilates) an electron at position ${\bf r}$, and $\alpha$ is the fine-structure constant.
The eigenstates $|\Psi_i\rangle$ of the molecular Hamiltonian $H_0$ are defined by the equation
\begin{equation}
H_0 | \Psi_i\rangle = E_i | \Psi_i\rangle.
\end{equation}
At time $t=t_0$, the vector potential is $\textbf{A}({\bf r},t_0)=0$ and
the molecule is in its ground state
\begin{equation}
|\Psi(t_0)\rangle = | \Psi_0 \rangle.
\end{equation}

In the presence of the electromagnetic field,
the total wave function in the interaction picture evolves in time according to the perturbation series (we employ atomic units, where $\hbar$ is set to $1$)
\begin{gather}
|\Psi(t) \rangle = | \Psi_0\rangle 
- i \sum_i  |\Psi_i\rangle  \int_{t_0}^{t} dt_1 e^{-\frac{\Gamma_i}{2} (t-t_1)}
  e^{i E_i t_1} e^{-i E_0 t_1}
\langle \Psi_i |  H_I(t_1) | \Psi_0\rangle  
\nonumber \\
+ (-i)^2 
\sum_{j,i} 
|\Psi_j\rangle
\int_{t_0}^{t} dt_2  
\int_{t_0}^{t_2} dt_1
e^{-\frac{\Gamma_i}{2} (t_2-t_1)} 
e^{i E_j t_2}
e^{-i E_i (t_2-t_1)}
e^{-i E_0 t_1}
\nonumber \\
\times \langle \Psi_j |   H_I(t_2) | \Psi_i \rangle 
\langle \Psi_i |  H_I(t_1)  | \Psi_0\rangle  
+ \dots , \label{eq:perturb}
\end{gather}
where we included an exponential decay with decay rate $\Gamma_i$
in the evolution of the states $i$ 
to incorporate the finite lifetime of the core-ionized states.
Because we are only addressing photoelectrons originating from the molecule before
the core-hole has collapsed via Auger decay, the decay of the state $j$ turns out to be irrelevant.
In Eq.~(\ref{eq:perturb}) the dots indicate higher order terms of the series, which are neglected in the following.

We want to monitor the dynamics of the single-core-hole wave packet
via photoelectrons resulting from double-core-ionization.
Our observable,
\begin{equation}
O_{q,q'}=\sum^{h<h'}_{h,h'} | \Psi_{h,h'}^{q,q' } \rangle \langle \Psi_{h,h'}^{q,q'} |,
\end{equation}
therefore involves the projection onto the two-particle-two-core-hole configurations described by the one-particle quantum numbers 
$q,q'$ for the photoelectrons in the unbound continuum and $h,h'$ for the two core holes, respectively. 
In our observable, we take the trace over all double-core-hole configurations $h,h'$.
Employing the time evolution of the wave packet in Eq.~(\ref{eq:perturb}),
the probability to observe a particular photoelectron pair with quantum numbers $q$ and $q'$ is accordingly
\begin{gather}
  P_{q,q'} = \langle \Psi(t\to \infty) | O_{q,q'}|\Psi(t\to \infty) \rangle =
\sum^{h<h'}_{h,h'} |\langle \Psi_{h,h'}^{q,q'} |\Psi(t\to \infty) \rangle |^2\\
= \alpha^4 \sum^{h<h'}_{h,h'}  \Big| 
\sum_{j,i} 
\langle \Psi_{h,h'}^{q,q'} |\Psi_j\rangle
\int_{t_0}^{\infty} dt_2  \int_{t_0}^{t_2} dt_1 
e^{i E_j t_2}
e^{-i E_i (t_2-t_1)}
e^{-\Gamma_i/2 (t_2-t_1)}
e^{-i E_0 t_1}
\nonumber \\
\times \langle \Psi_j |   H_I(t_2) | \Psi_i \rangle 
\langle \Psi_i |  H_I(t_1)| \Psi_0\rangle  
\Big|^2.\label{eq:pqq}
\end{gather}
In Eq.~(\ref{eq:pqq}) we have neglected {\it direct} one-photon double-core-ionization (i.e., K-shell hyper-satellites), 
which has for K-shell ionization a negligible cross section\cite{goldsztejn_doublecorehole_2016,marchenko_ultrafast_2018a}.

We consider an x-ray pulse consisting of two sub-pulses, each of which has a Gaussian envelope.
The first pulse is centered at time $t=0$ and the second is centered at time $t=t_d$.
The  vector potential ${\bf A}({\bf r},t)$ of such a pulse is
\begin{eqnarray}
{\bf A}({\bf r},t)&=&{\bf \hat{\epsilon}} \text{Re} \{ \left(  A_1(t)+A_2(t) \right)  \}\\
A_1(t)&=& A_1 e^{-\frac{t^2}{2\sigma^2}} e^{i {\bf k}_1 \cdot {\bf r}} e^{-i \omega_1 t}\\
A_2(t)&=& A_2 e^{-\frac{(t-t_d)^2}{2\sigma^2}} e^{i {\bf k}_2 \cdot {\bf r}} e^{-i \omega_2 t},
\end{eqnarray}
where ${\bf \hat{\epsilon}}$ is a unit vector along the polarization direction (both sub-pulses have the same polarization), 
${\bf k}_1$ and ${\bf k}_2$ are the photon momentum vectors of the two pulses, respectively, $\sigma$ is the temporal width of each of the two sub-pulses, and $\omega_1=|{\bf k}_1 |/\alpha$ and $\omega_2=|{\bf k}_2 |/\alpha$
are the central photon energies of both sub-pulses.

Inserting this pulse into Eq.~(\ref{eq:pqq}) and
ignoring counter-rotating terms,
the probability to observe the photoelectron pair becomes

\begin{gather}
P_{q,q'} 
= \alpha^4 \sum^{h<h'}_{h,h'}  \Big| 
\sum_{i,j}
\langle \Psi_{h,h'}^{q,q'} |\Psi_j\rangle 
\nonumber \\
\Big(
\langle \Psi_j|T_{{\bf k}_1} | \Psi_i \rangle 
\langle \Psi_i |T_{{\bf k}_1} | \Psi_0\rangle   
A_1^2 I_{11}(E_i-E_0-\omega_1,E_j-E_i-\omega_1)\nonumber \\
+
\langle \Psi_j|T_{{\bf k}_2} | \Psi_i \rangle 
\langle \Psi_i |T_{{\bf k}_2} | \Psi_0\rangle   
A_2^2 I_{22}(E_i-E_0-\omega_2,E_j-E_i-\omega_2)\nonumber \\
+
\langle \Psi_j|T_{{\bf k}_2} | \Psi_i \rangle 
\langle \Psi_i |T_{{\bf k}_1} | \Psi_0\rangle   
A_1 A_2 I_{12}(E_i-E_0-\omega_1,E_j-E_i-\omega_2)\nonumber \\
+
\langle \Psi_j|T_{{\bf k}_1} | \Psi_i \rangle 
\langle \Psi_i |T_{{\bf k}_2} | \Psi_0\rangle   
A_1 A_2 I_{21}(E_i-E_0-\omega_2,E_j-E_i-\omega_1)
\Big)
\Big|^2, \label{eq:ppn2}
\end{gather}
where 
\begin{equation}
T_{\bf k}=\int d^3 r\, \psi^\dagger({\bf r})  \, {\bf \hat{\epsilon}} \cdot {\bf p} \, e^{i {\bf k}\cdot  {\bf r}} \, \psi({\bf r})
\end{equation}
and the following abbreviations for the time integrals have been used:
\begin{eqnarray}
  I_{11}(\Delta E_1,\Delta E_2)&=&\int_{-\infty}^{\infty} dt_2
  e^{-\frac{t_2^2}{2\sigma^2}}
  e^{- \frac{\Gamma}{2} t_2}
  e^{i \Delta E_2 t_2}
  \int_{-\infty}^{t_2} dt_1
  e^{-\frac{t_1^2}{2\sigma^2}}
  e^{\frac{\Gamma}{2} t_1}    
  e^{i \Delta E_1 t_1}\\
 I_{22}(\Delta E_1,\Delta E_2)&=&\int_{-\infty}^{\infty} dt_2
 e^{-\frac{(t_2-t_d)^2}{2\sigma^2}}
 e^{ - \frac{\Gamma}{2} t_2}
 e^{i \Delta E_2 t_2}
 \int_{-\infty}^{t_2} dt_1
 e^{-\frac{(t_1-t_d)^2}{2\sigma^2}}
 e^{\frac{\Gamma}{2} t_1}    
 e^{i \Delta E_1 t_1}\\
 I_{12}(\Delta E_1,\Delta E_2)&=&
 \int_{-\infty}^{\infty} dt_2
 e^{-\frac{(t_2-t_d)^2}{2\sigma^2}}
 e^{-\frac{\Gamma}{2} t_2}
 e^{i \Delta E_2 t_2}
 \int_{-\infty}^{t_2} dt_1
 e^{-\frac{t_1^2}{2\sigma^2}}
 e^{\frac{\Gamma}{2} t_1}    
 e^{i \Delta E_1 t_1}\label{eq:timeintegral}\\
 I_{21}(\Delta E_1,\Delta E_2)&=&
 \int_{-\infty}^{\infty} dt_2
 e^{-\frac{t_2^2}{2\sigma^2}}
 e^{-\frac{\Gamma}{2} t_2}
 e^{i \Delta E_2 t_2}
 \int_{-\infty}^{t_2} dt_1
 e^{-\frac{(t_1-t_d)^2}{2\sigma^2}}
 e^{\frac{\Gamma}{2} t_1}    
 e^{i \Delta E_1 t_1}.
\end{eqnarray}

Equation~(\ref{eq:ppn2}) involves terms where both photoionizations happen in the first pulse 
(proportional to $A_1^2$),
both photoionizations happen in the second pulse 
(proportional to $A_2^2$),
or where the first photoionization happens in the first and the second photoionization in the second pulse
and the swapped term (proportional to $A_1 A_2$) .
The accompanying time integrals 
define the spectral positions and the line-shapes for the two corresponding photoelectrons.
Trivially, the time integrals where both photoionizations
occur in the same pulse show no dependence on the delay time $t_d$.

In the following, we neglect vibrational dynamics and assume a static molecular geometry.
As discussed in detail in Ref.~\onlinecite{ehara_symmetrydependent_2006}, the minimum of the 
core-ionized potential energy curves is shifted from the neutral ground-state potential energy curve by 2\% 
towards a shorter bond length, whereas the curvature stays almost the same.
The Franck-Condon ratio that indicates the amount of vibrational excitation upon core ionization is below 15\%.
Thus, core-ionizing the \ce{N2} molecule induces little vibrational excitation, and we can therefore assume
that vibrational effects can be neglected for the further discussion.
Accordingly, we can write the states $|\Psi_i\rangle$ as 
many-electron wave functions that are composed of an $N-1$ electron bound, core-ionized eigenstate and an unbound photoelectron, $|\Psi_i\rangle = |\Psi_{n}^{q_1}\rangle$. 
In a similar manner, the states $|\Psi_j\rangle$ are assumed to be anti-symmetrized products of 
$N-2$ bound-electron wave functions describing an eigenstate with a double-core hole and two photoelectrons, 
$|\Psi_j\rangle = |\Psi_{m}^{q_1 q_2}\rangle$.
We neglect any interaction of the two photoelectrons such that
the energy of the states $E_i$ and $E_j$ can be written as the energy 
of the respective single- or double-core-ionized state plus 
the asymptotic kinetic energy of the photoelectrons,
\begin{gather}
E_i=E_n+\epsilon_{q_1}, \quad E_j=E_m+\epsilon_{q_1}+\epsilon_{q_2}.
\end{gather}

The core-ionization potential of single-core-hole states $E_m-E_n$ is significantly higher than the core-ionization 
potential of the neutral molecule $E_n-E_0$ due to the hole-hole interaction in the molecule.
This effect is particularly strong for single-site double-core-hole (SSDCH) states, 
where the binding energy for the second ionization is $\simeq \unit[80]{eV}$ higher than
the binding energy for the first core-ionization step\cite{tashiro_double_2012};
for two-site double-core-hole states (TSDCH), the second binding energy is $\simeq \unit[20]{eV}$ higher\cite{carravetta_symmetry_2013}.
In the following, we focus on observing specific photoelectron pairs $q_1,q_2$ that arise from 
core ionization with the first pulse and second core ionization with the second pulse.
We choose appropriate photon energies, $\omega_1$ and $\omega_2$, 
and we focus on specific photoelectron energies $\epsilon_{q_1}$ and $\epsilon_{q_2}$,
such that only photoelectrons that result from 
ionization of the neutral by the first pulse and ionization of the cation by the second pulse are filtered out.
In other words, we demand that all terms in Eq.~(\ref{eq:ppn2}) except the term proportional to 
$I_{12}(E_i-E_0-\omega_1,E_j-E_i-\omega_2)$ vanish.

The probability to observe such a pair of photoelectrons with distinct energy $\epsilon_{q_1} \ne \epsilon_{q_2}$ is
\begin{gather}
P_{q_1,q_2} 
= \alpha^4 |A_1|^2 |A_2|^2 \sum_{m}^{\text{DCH}}  \Big| 
\sum_{n}^{\text{SCH}}  
\langle \Psi_m^{q_1 q_2}|T_{{\bf k}_2} | \Psi_n^{q_1} \rangle 
\langle \Psi_n^{q_1} |T_{{\bf k}_1} | \Psi_0\rangle   \nonumber \\
\times I_{12}(\epsilon_{q_1}+E_n-E_0-\omega_1,\epsilon_{q_2}+E_m-E_n-\omega_2)
\Big|^2, \label{eq:ppn3}
\end{gather}
where the sum $m$ and $n$ runs over double-core-hole (DCH) and single-core-hole (SCH) eigenstates.
The remaining time integral $I_{12}(\epsilon_{q_1}+E_n-E_0-\omega_1,\epsilon_{q_2}+E_m-E_n-\omega_2)$ 
in Eq.~(\ref{eq:ppn3}) can be carried out 
under the assumption that the pulses do not overlap ($t_d \gg \sigma$)
and yields
\begin{eqnarray}
I_{12}(\Delta E_1,\Delta E_2)
&\simeq&
2\pi \sigma^2
e^{\frac{-\Delta E_1^2}{2/\sigma^2}}
e^{\frac{-\Delta E_2^2}{2/\sigma^2}}
e^{-\frac{\Gamma}{2} t_d} e^{ i \Delta E_2 t_d}\\
&=&I(\Delta E_1) I(\Delta E_2) e^{-\frac{\Gamma}{2} t_d} e^{i \Delta E_2 t_d}, \label{eq:timeintegral_simplified}
\end{eqnarray}
where we have ignored finite lifetime effects over the finite bandwidth of the pulse, ($\Gamma\sigma \simeq 0$).
The case of finite overlap between two pulses is discussed in Sec.~\ref{sec:finite_overlap}.

Given the fact that the spectral bandwidth of the x-ray pulses, $1/\sigma$,
is supposed to be much larger than the energy separation between the core-hole binding energies,
$\Delta \epsilon_{gu}\simeq \unit[100]{meV}$, 
the time integral $I(\Delta E)$ does not depend on the specific single core-hole state,
so we can further approximate
\begin{equation}
I(\epsilon_{q_1}+E_n-E_0-\omega_1) \simeq I(\epsilon_{q_1}-\epsilon_{K}-\omega_1 ) 
\end{equation}
where $\epsilon_K$ is the averaged core-hole binding energy.
For double-core-hole states, we have two nearly degenerate pairs of states:
The SSDCH states are characterized by the two configurations
$1\sigma_g^{-2}+1\sigma_u^{-2}$ and $1\sigma_g^{-1}1\sigma_u^{-1}$ (singlet),
while the TSDCH states are characterized by the two configurations
$1\sigma_g^{-2}-1\sigma_u^{-2}$ and $1\sigma_g^{-1}1\sigma_u^{-1}$ (triplet).
As with the single-core-hole states, we can neglect the energy difference between 
the specific double-core-hole state pairs in the integral $I(\Delta E)$, thus we approximate 
\begin{equation}
I(\epsilon_{q_2}+E_m-E_n-\omega_2) \simeq I(\epsilon_{q_2}-\epsilon_{KK}-\omega_2 ),
\end{equation}
where $\epsilon_{KK}$ is the appropriate double-core-hole binding energy (either SSDCH or TSDCH).
With this approximation, we obtain
\begin{gather}
P_{q_1,q_2} 
= \alpha^4 |A_1|^2 |A_2|^2 
I^2(\epsilon_{q_1}-\epsilon_K-\omega_1)
I^2(\epsilon_{q_2}-\epsilon_{KK}-\omega_2)
e^{-\Gamma t_d}\nonumber \\
\times \sum_{m}^{\text{DCH}}  
\Big| 
\sum_{n}^{\text{SCH}}  
 e^{-i E_n t_d}
\langle \Psi_m^{q_2}|T_{{\bf k}_2}| \Psi_n \rangle 
\langle \Psi_n^{q_1} |T_{{\bf k}_1}| \Psi_0\rangle  \Big|^2, \label{eq:ppn4}
\end{gather}
where the sum $m$ runs either over SSDCH or TSDCH states.

To evaluate the interaction matrix elements, we now employ a basis 
for the continuum electrons defined by the asymptotic momentum vector ${\bf q}$ and assume
simple spin- and symmetry-adapted configurations for the core-hole states.
Furthermore, we employ the dipole approximation, i.e.,
\begin{equation}
T_{\bf k}=\int d^3 r\, \psi^\dagger({\bf r})  \, {\bf \hat{\epsilon}} \cdot {\bf p} \, e^{i {\bf k}\cdot  {\bf r}} \, \psi({\bf r}) \simeq \int d^3 r\, \psi^\dagger({\bf r})  \, {\bf \hat{\epsilon}} \cdot {\bf p} \, \psi({\bf r}) = T_{\bf 0}.
\end{equation}
Carrying out the sums over DCH and SCH states in these minimal spin- and symmetry-adapted configurations and 
performing a summation over the spins of the two photoelectrons yields for the SSDCH states
\begin{gather}
P^{\text{SSDCH}}_{{\bf q_1},{\bf q_2}} 
= \frac{\alpha^4}{4} |A_1|^2 |A_2|^2 
I^2(\epsilon_{q_1}-\epsilon_K-\omega_1)
I^2(\epsilon_{q_2}-\epsilon^{\text{SSDCH}}_{KK}-\omega_2)e^{-\Gamma t_d}\nonumber \\
\times \Bigg(
|\mu_{{\bf q}_2,1 \sigma_g }|^2
|\mu_{{\bf q}_1,1 \sigma_g }|^2
+
|\mu_{{\bf q}_2,1 \sigma_u }|^2
|\mu_{{\bf q}_1,1 \sigma_g }|^2\nonumber \\
+
|\mu_{{\bf q}_2,1 \sigma_u }|^2
|\mu_{{\bf q}_1,1 \sigma_u }|^2
+
|\mu_{{\bf q}_2,1 \sigma_g }|^2
|\mu_{{\bf q}_1,1 \sigma_u }|^2\nonumber \\
+
4\text{Re}\{
 e^{i \Delta \epsilon_{gu} t_d}
\mu_{{\bf q}_1,1 \sigma_g } 
\mu_{{\bf q}_1,1 \sigma_u }^*
\}
\text{Re} \{
\mu_{{\bf q}_2,1 \sigma_g }
\mu_{{\bf q}_2,1 \sigma_u }^*
\}
\Bigg)\label{eq:SSDCHppn5}
\end{gather}
and for the TSDCH states
\begin{gather}
P^{\text{TSDCH}}_{{\bf q_1},{\bf q_2}} 
= \alpha^4 |A_1|^2 |A_2|^2 
I^2(\epsilon_{q_1}-\epsilon_K-\omega_1)
I^2(\epsilon_{q_2}-\epsilon^{\text{TSDCH}}_{KK}-\omega_2)e^{-\Gamma t_d}
\nonumber \\
\times \Bigg(
\frac{1}{4}|\mu_{{\bf q}_2,1 \sigma_g }|^2
|\mu_{{\bf q}_1,1 \sigma_g }|^2
+
\frac{1}{4}|\mu_{{\bf q}_2,1 \sigma_u }|^2
|\mu_{{\bf q}_1,1 \sigma_u }|^2 \nonumber \\
+
\frac{3}{4}|\mu_{{\bf q}_2,1 \sigma_u }|^2
|\mu_{{\bf q}_1,1 \sigma_g }|^2
+
\frac{3}{4}
|\mu_{{\bf q}_2,1 \sigma_g }|^2
|\mu_{{\bf q}_1,1 \sigma_u }|^2\nonumber \\
-\text{Re}\{
e^{i \Delta \epsilon_{gu} t_d}
\mu_{{\bf q}_1,1 \sigma_g } 
\mu_{{\bf q}_1,1 \sigma_u }^*
\}
\text{Re} \{
\mu_{{\bf q}_2,1 \sigma_g }
\mu_{{\bf q}_2,1 \sigma_u }^*
\}\nonumber  \\
-
 \text{Re} \{
e^{i \Delta \epsilon_{gu} t_d}
\mu_{{\bf q}_2,1 \sigma_u }
\mu_{{\bf q}_2,1 \sigma_g }^*
\mu_{{\bf q}_1,1 \sigma_g }
\mu_{{\bf q}_1,1 \sigma_u }^*
\}
\Bigg) \label{eq:TSDCHppn5}.
\end{gather}

In Eq.~(\ref{eq:SSDCHppn5}) and Eq.~(\ref{eq:TSDCHppn5}), we have used the abbreviation
\begin{equation}
\mu_{{\bf q},h}=
{\bf \hat{\epsilon}} \cdot \langle {{\bf q}} | {\bf p}   | h \rangle \label{dipole}
\end{equation}
which is the momentum-form dipole matrix element along the polarization direction $\hat{\epsilon}$
evaluated for hole $h$ and a photoelectron with asymptotic momentum ${\bf q}$.

The dipole matrix elements can be written via atomic orbitals on the left and right atoms, $1sL$ and $1sR$.
Employing plane waves for the photoelectrons and neglecting atomic overlap contributions, one obtains
\begin{eqnarray}
\mu_{{\bf q},1\sigma g}&=\frac{1}{\sqrt{2}} \left( \mu_{{\bf q},1sL}+\mu_{{\bf q},1sR} \right)&=\sqrt{2} \mu_{{\bf q},1s}\cos( {\bf q}\cdot {\bf d}/2  )\\
\mu_{{\bf q},1\sigma u}&=\frac{1}{\sqrt{2}} \left( \mu_{{\bf q},1sL}-\mu_{{\bf q},1sR} \right)&=i \sqrt{2} \mu_{{\bf q},1s} \sin( {\bf q}\cdot {\bf d}/2),
\end{eqnarray}
where ${\bf d}$ is the vector describing the distance between the two atoms and 
\begin{equation}
\mu_{{\bf q},1s}= \hat{\epsilon} \cdot \langle {\bf q} | {\bf p} | 1s \rangle
\end{equation}
is the atomic transition matrix element.
Employing these dipole matrix elements, we get explicit analytical expressions
for the angular dependence.
Because the phase relation between ionization from $\sigma_g$ and $\sigma_u$ is such that $\text{Re} \{ \mu_{{\bf q}_2,1 \sigma_g }\mu_{{\bf q}_2,1 \sigma_u }^*\}=0$, any time-delay dependence vanishes (except for the exponential decay) for the SSDCH,
\begin{gather}
P^{SSDCH}_{{\bf q_1},{\bf q_2}} 
=\alpha^4 |A_1|^2 |A_2|^2 
I^2(\epsilon_{q_1}-\epsilon_K-\omega_1)
I^2(\epsilon_{q_2}-\epsilon_{KK}-\omega_2)e^{-\Gamma t_d}
|\mu_{{\bf q}_1,1s}|^2
|\mu_{{\bf q}_2,1s}|^2.
\end{gather}
For the TSDCH, the probability to detect the two photoeelectrons with momenta ${\bf q}_1$ and ${\bf q}_2$ reads
\begin{gather}
P^{TSDCH}_{{\bf q_1},{\bf q_2}} 
= \alpha^4 |A_1|^2 |A_2|^2 
I^2(\epsilon_{q_1}-\epsilon_K-\omega_1)
I^2(\epsilon_{q_2}-\epsilon_{KK}-\omega_2)e^{-\Gamma t_d}
|\mu_{{\bf q}_1,1s}|^2
|\mu_{{\bf q}_2,1s}|^2  \nonumber
\\
\times \Bigg(
2-\cos({\bf q}_1 \cdot {\bf d})\cos({\bf q}_2 \cdot {\bf d})
-\cos (\Delta \epsilon_{gu} t_d)
\sin({\bf q}_1 \cdot {\bf d})
\sin({\bf q}_2 \cdot {\bf d})
\Bigg). \label{eq:TSDCH}
\end{gather}
Remarkably, for the TSDCH, the angular distribution of the two photoelectrons depends on the delay time 
as $\cos(\Delta \epsilon_{gu} t_d)$, i.e., it oscillates with the position of the core hole.

It is instructive to compare this result with the core-hole wave-packet dynamics. 
The nonstationary core-hole wave packet that is induced by the photoionization emitting
a photoelectron into direction ${\bf q}_1$ is 
\begin{gather}
|\Psi_{\text{core hole}} (t) \rangle  \propto  e^{i \epsilon_{1\sigma g} t} \cos ({\bf q }_1 \cdot {\bf d}/2) |\Psi_{1\sigma_g^{-1}} \rangle + i e^{i \epsilon_{1\sigma u} t} \sin ({\bf q }_1\cdot  {\bf d}/2) |\Psi_{1\sigma_u^{-1}}\rangle, \label{eq:corehole_wave_packet}
\end{gather}
and gives rise to the hole-density oscillations
\begin{gather}
\rho(t)  \sim 
(\rho_{1sL^{-1}} + \rho_{1sR^{-1}} ) 
-\sin ({\bf q }_1 \cdot {\bf d})
\sin(\Delta \epsilon_{gu} t)
(\rho_{1sL^{-1}} - \rho_{1sR^{-1}}  ). \label{eq:hole_density}
\end{gather}
As can be seen, for photoelectrons emitted into directions perpendicular to the molecular axis (${\bf q }_1 \cdot {\bf d}=0$) the 
wave function of the molecular cation is a pure $1\sigma_g^{-1}$ core-hole state. 
At this angle, the core hole is fully delocalized over both atoms and does not oscillate. 
Accordingly, no delay time dependence is seen.
For other angles, we obtain a linear combination of $1\sigma_g$ and $1\sigma_u$ holes that 
gives rise to an initially delocalized hole that oscillates 
with an amplitude $\sin(q_1 d \cos(\theta_1))$, where $\theta_1$ is the emission angle of the first photoelectron with respect to the molecular axis.
A maximal localization of the hole on either the left or the right atom can be achieved at angles 
\begin{equation}
  \theta_1=0 \qquad \text{and} \qquad \theta_1=\text{acos}(\pm \pi/(2 {q}_1 d)). \label{eq:theta1max}
\end{equation}
Accordingly, the oscillations with delay time according to Eq.~(\ref{eq:TSDCH}) are largest for these angles.
For the second photoelectron, 
the time-delay-dependent contribution is proportional to $\sin(q_2 d \cos(\theta_2))$, where $\theta_2$ is the emission direction with respect to the molecular axis. 
The largest oscillations with delay time $t_d$ are expected for
\begin{gather}
  \theta_2=0 \qquad \text{and} \qquad
  \theta_2=\text{acos}(\pm \pi/(2 {q}_2 d)) \label{eq:thetamax2},
\end{gather}
that is, when the probe pulse ionizes from the molecular ion a maximally localized core electron.

Equation~(\ref{eq:TSDCH}) expresses the photoelectron angular distribution with respect to the molecular axis 
via the scalar product ${\bf d} \cdot {\bf q}=d q \cos(\theta)$.
The factor containing the atomic transition matrix element, $|\mu_{{\bf q}_1,1s}|^2 |\mu_{{\bf q}_2,1s}|^2$, 
involes an additional dependency with respect to the polarization axis.
To reduce the complexity of the angular distribution, we are now discussing two strategies:
The angular distribution of the photoelectrons with respect to the molecular axis after averaging over all possible orientations of the polarization axis (in the following referred to as molecular-frame angular distribution), 
or, the angular distribution of the photoelectron with respect to the polarization axis after averaging over all possible orientations of the molecular axis (lab frame).

When calculating the molecular-frame photoelectron angular distribution, we integrate over the azimuthal angles $\phi_1$ and $\phi_2$ of the photoelectrons with respect to the molecular axis.
The polarization-dependent factor in Eq.~(\ref{eq:TSDCH}) becomes
\begin{gather}
\int_0^{2\pi} d \phi_1
\int_0^{2\pi} d \phi_2
\frac{1}{4\pi}
  \int d \Omega_{\hat{\epsilon}}
  |\hat{\epsilon} \cdot \langle {\bf q}_1 | {\bf p} | 1s \rangle |^2
  |\hat{\epsilon} \cdot \langle {\bf q}_2 | {\bf p} | 1s \rangle |^2 \nonumber \\
=\frac{4\pi^2}{15}|\langle {\bf q}_1 | {\bf p} | 1s \rangle|^2 |\langle {\bf q}_2 | {\bf p} | 1s \rangle|^2\left(1+2\cos^2(\theta_1)\cos^2(\theta_2)+\sin^2(\theta_1)\sin^2(\theta_2) \right)
\label{eq:pol_avg}.
\end{gather}

For the lab-frame photoelectron angular distribution we obtain
\begin{gather}
\frac{1}{4\pi}\int d\Omega_d P^{TSDCH}_{{\bf q_1},{\bf q_2}} =
\alpha^4 |A_1|^2 |A_2|^2 
I^2(\epsilon_{q_1}-\epsilon_K-\omega_1)
I^2(\epsilon_{q_2}-\epsilon_{KK}-\omega_2)e^{-\Gamma t_d}
|\mu_{{\bf q}_1,1s}|^2
|\mu_{{\bf q}_2,1s}|^2 \nonumber
\\
\times \Bigg(
2 - 
A_{+} -  \cos (\Delta \epsilon_{gu} t_d) A_{-}
\Bigg), \label{eq:TSDCH_d_averaged}
\end{gather}
with 
\begin{equation}
A_{\pm}=\frac{\sin \left(d|{\bf q}_1-{\bf q}_2| \right)}{2d|{\bf q}_1-{\bf q}_2|}
\pm \frac{\sin \left(d|{\bf q}_1+{\bf q}_2| \right)}{2 d|{\bf q}_1+{\bf q}_2|} 
\end{equation}
and 
\begin{gather}
|{\bf q}_1\pm{\bf q}_2|=\sqrt{q_1^2+q_2^2\pm2q_1q_2\cos(\Theta_1)\cos(\Theta_2)+\cos(\Phi_1-\Phi_2)\sin(\Theta_1)\sin(\Theta_2)},
\end{gather}
where $\Theta_1,\Theta_2$ are the two photoelectron emission angles with respect to the polarization axis and 
$\Phi_1,\Phi_2$ the corresponding azimuthal angles.

\section{Numerical Results \label{sec:numeric}}

To further discuss the properties of the angular dependency of the two photoelectrons, we now specify explicit x-ray pulse parameters.
The binding energy for the first core ionization in \ce{N2} is 
$\epsilon_K\simeq \unit[410]{eV}$\cite{carravetta_symmetry_2013}, 
the binding energy for the second core ionization (assuming two-site double-core holes) is $\epsilon_{KK}\simeq \unit[425]{eV}$\cite{carravetta_symmetry_2013}.
We choose the photon energies to be $\omega_1=\unit[430]{eV}$ and $\omega_2=\unit[440]{eV}$.
With this choice, the resulting photoelectron energies are $|{\bf q_1}|^2/2=\omega_1-\epsilon_K \simeq \unit[20]{eV}$, $|{\bf q_2}|^2/2=\omega_2-\epsilon_{KK}=\unit[15]{eV}$.
For the molecular parameters, we employ the values $d=\unit[2.078]{a.u.}$, $\Delta \epsilon_{gu}=\unit[100]{meV}$, and $1/\Gamma\simeq \unit[7]{fs}$.

\subsection{No Pulse Overlap \label{sec:no_overlap}}
To discuss the angular correlation of the two photoelectrons with respect to the molecular axis, we first discuss results 
assuming no temporal overlap of the two x-ray pulses.
The effects of finite pulse overlap will be discussed in Sec.~\ref{sec:finite_overlap}.

\begin{figure}
  \includegraphics[width=0.8\textwidth]{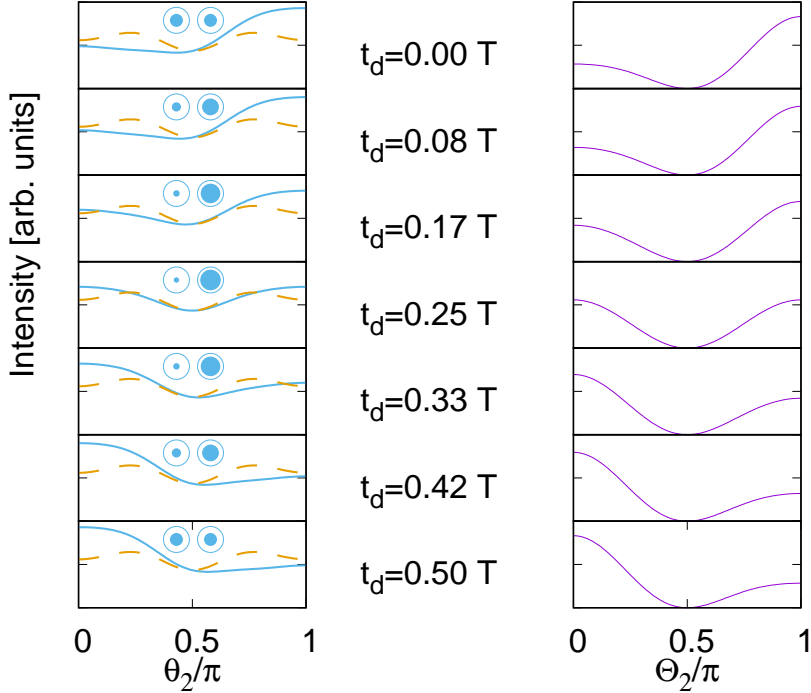}
  \caption{Intensity of the second photoelectron as a function of emission angle for selected delay times $t_d$ and fixed emission angle 
    of the first photoelectron and averaged over the orientation of the molecule with respect to the polarization axis.         
    The left panel shows the angular dependence in the molecular frame. The solid line shows the angular dependence when the first photoelectron has been emitted along the molecular axis ($\theta_1=0$). The dashed line shows the constant angular dependence when the first electron has been emitted perpendicular to the molecular axis ($\theta_1=\pi/2$).
    The right panel shows the angular dependence in the lab frame, where the first photoelectron has been emitted along the polarization axis ($\Theta_1=0$).
    $T$ is the core-hole oscillation period of \unit[44.2]{fs}.
    The oscillation of the hole density from the left to the 
    right atom is sketched in each sub-plot with the filled circle representing presence of the hole.
    \label{fig:compare_orient_avg}
  }
\end{figure}

We first compare the two strategies discussed before.
Figure~\ref{fig:compare_orient_avg} shows 
the molecular-frame angular distribution in the left panel and 
the lab-frame angular distribution in the right panel.

For the molecular frame (left panel), we show two cases: 
First, we assume that the first photoelectron has been emitted along
the molecular axis ($\theta_1=0$) for which a maximal mixing of $\sigma_g$ and $\sigma_u$ core holes can be found (see Eq. 37).
As a reference, we also compare with the case where the first photoelectron is emitted perpendicular to the molecular axis and no oscillation of the angular pattern appears (dashed line in Fig.~\ref{fig:compare_orient_avg}).
For all time steps the photoelectron spectra are normalized such that the exponential decay cancels out.
We note however, that the detected signal drops due to the exponential decay of the core hole state.
For the molecular frame (left panel), the hole density on the two atoms for the respective delay time 
as given by Eq.~(\ref{eq:hole_density}) is indicated by the small sketch of the molecule in each plot.
Such a direct interpretation of hole density and emission angle is not possible for the lab frame approach.

In both panels, the angular dependency looks remarkably similar, although one has to keep in mind that the two angles $\theta$ and $\Theta$ 
have a different meaning.
Both approaches, the photoelectron angular distribution in the lab frame and in the molecular frame, show a pronounced oscillation of the emission direction of the second photoelectron
as a function of delay time.
In the molecular-frame (left panel), one can see that there is a significant 
left-right asymmetry in the emission of the second photoelectron at $t_d=0$, where the hole is delocalized over both atoms. 
Initially, the second photoelectron tends to be emitted into the opposite direction ($\theta_2=\pi$) relative to the first photoelectron.
The emission is symmetric with respect to left and right for $t_d=T/4$, where the hole is almost localized on one atom.
Finally, at $t_d=T/2$ the second photoelectron tends to be emitted in the same direction as the first photoelectron ($\theta_2=0$).
In the lab frame (right panel) a similar oscillation is seen as in the molecular frame.
Because in the lab frame the emission is modulated via
the $\cos^2(\Theta_2)$ factor, the signal at $\Theta_2=\pi/2$ is always zero.
In contrast, in the molecular frame, there is a background for all directions. 

As we have seen, both of the discussed strategies give remarkably strong oscillations of the photoelectron angular distribution.
Because the angle $\theta_1$ determines 
the amplitude and phase, and thus the initial state, 
of the core-hole wave packet, (as expressed in Eq.~(\ref{eq:hole_density})), 
the photoelectron angular-distribution in the molecular frame is the more natural choice.
In the experiment, this can be achieved via coincident detection of the two photoelectron with one of the ions.
The experimental feasibility is discussed in Sec.~\ref{sec:realization}.

\begin{figure}
\subfigure[$t_d=0$]{
  \includegraphics[width=0.45\textwidth]{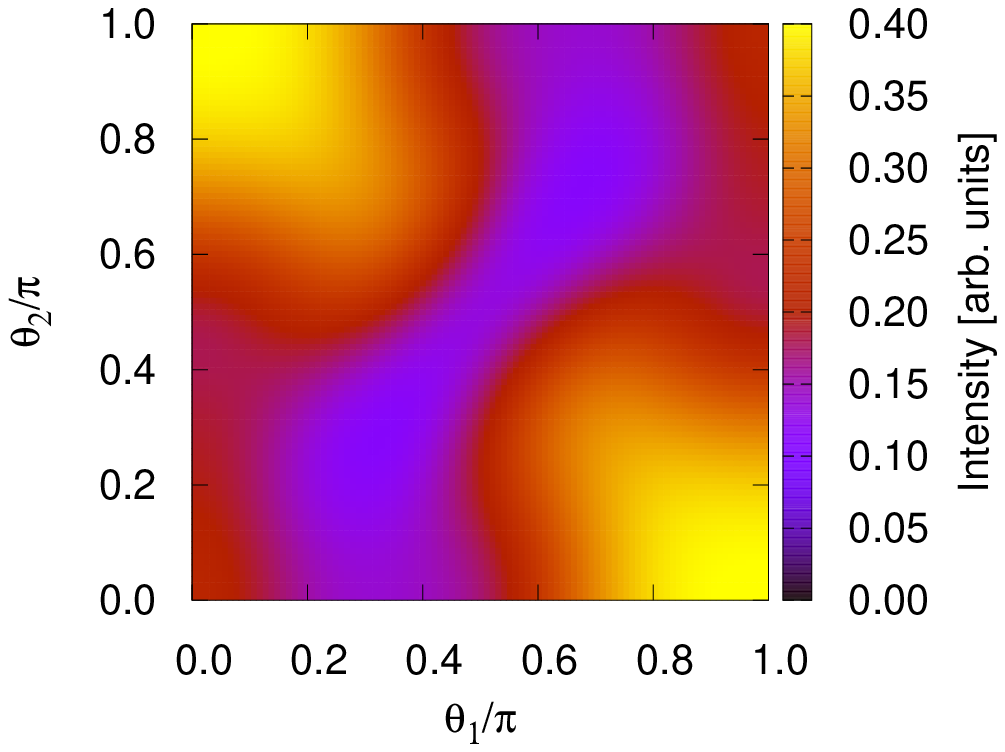}}
\subfigure[$t_d=\frac{1}{4} T$]{
  \includegraphics[width=0.45\textwidth]{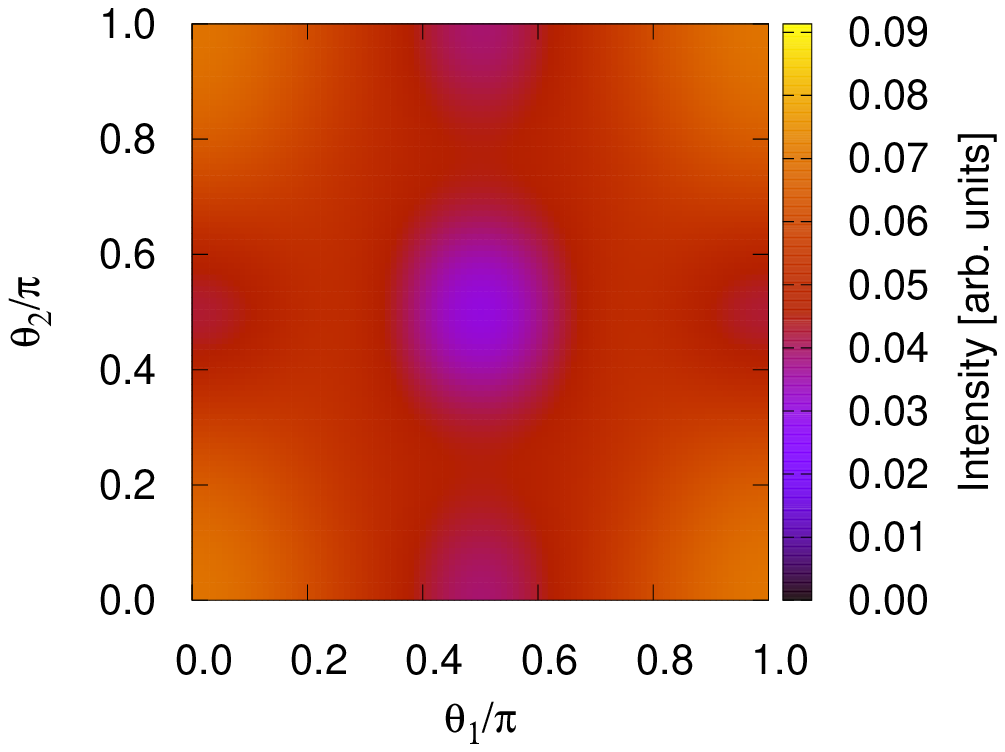}}
\subfigure[$t_d=\frac{1}{2} T$]{
  \includegraphics[width=0.45\textwidth]{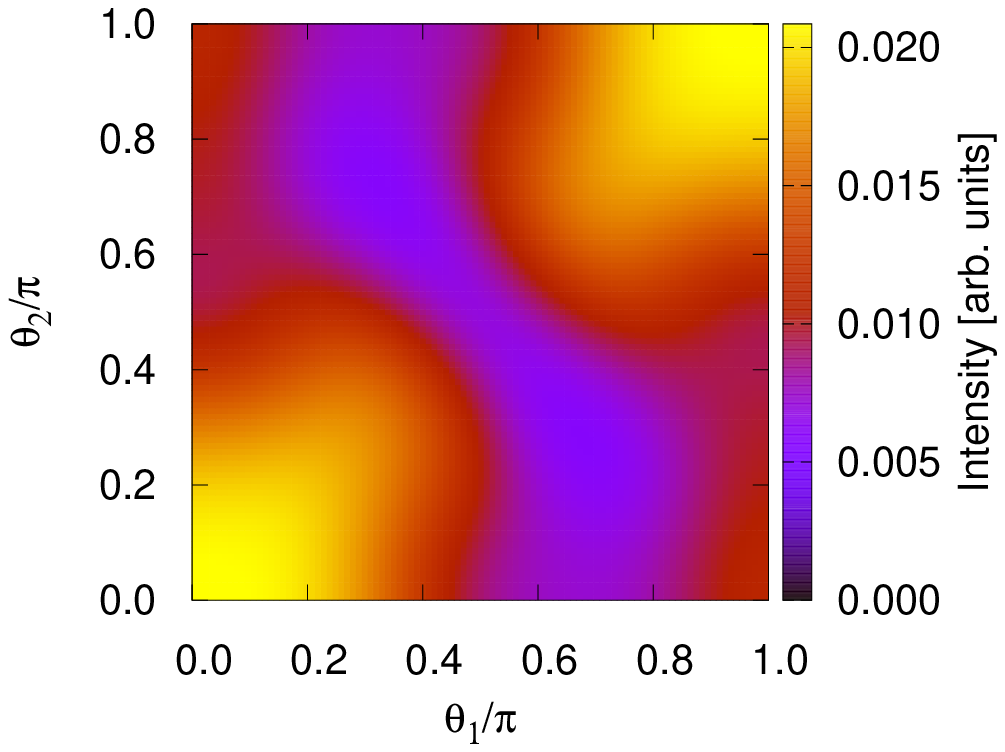}}
\subfigure[$t_d=T$]{
  \includegraphics[width=0.45\textwidth]{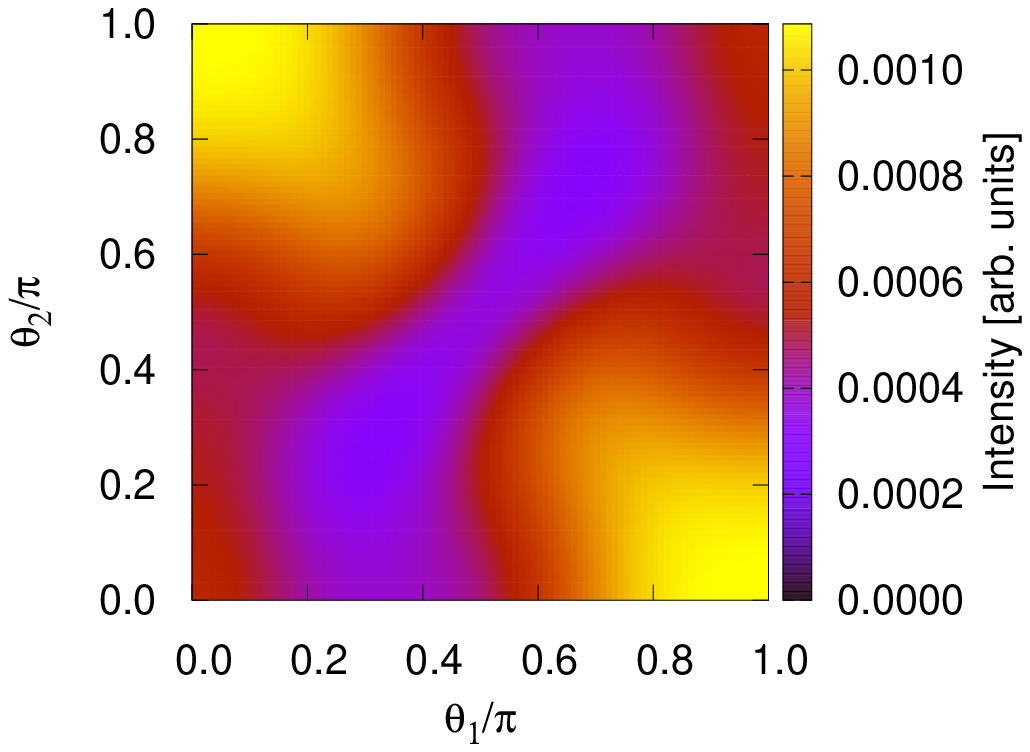}}
\caption{Intensity of two photoelectrons stemming from core-shell ionization and subsequent double-core-shell ionization
  integrated over azimuthal angles $\phi_1$ and $\phi_2$ and shown
  as a function of angles $\theta_1$ and $\theta_2$ to the molecular axis 
  for selected delay times ( (a)~$t_d=0$, (b)~$t_d=\frac{1}{4} T$, (c)~$t_d=\frac{1}{2} T$, and (d)~$t_d=T$).
  An average over the orientation of the molecular axis with respect to the polarization axis is assumed.
  $T$ is the core-hole oscillation period of $\unit[41.4]{fs}$.
  Each subfigure has an adapted color palette due to the overall exponential decay of the signal.
\label{fig:angle12}
}
\end{figure}
Figure~\ref{fig:angle12} shows the intensity as a function of both angles $\theta_1$ and $\theta_2$ for selected delay times.
Note that the overall exponential decay is reflected here in the adapted color palette
in each sub-figure.
As can be seen, the photoelectrons tend to be emitted into opposite hemispheres of the molecule initially, as indicated by the
low values on the diagonal of the angular-angular plot and the larger values at 
$\theta_1=\pi$, $\theta_2=0$ and $\theta_1=0$, $\theta_2=\pi$.
At $t_d=T/4 \simeq \unit[10]{fs}$, there is an 
equal contribution for the two photoelectrons to be emitted into the same direction ($\theta_1=\theta_2=0$, $\theta_1=\theta_2=\pi$) and into the opposite directions ($\theta_1=0,\theta_2=\pi$ and $\theta_1=\pi, \theta_2=0$).
After half a period at $t_d=T/2 \simeq \unit[21]{fs}$ the photoelectrons tend to be emitted into
same hemispheres of the molecule. 

\begin{figure}
\includegraphics[width=0.8\textwidth]{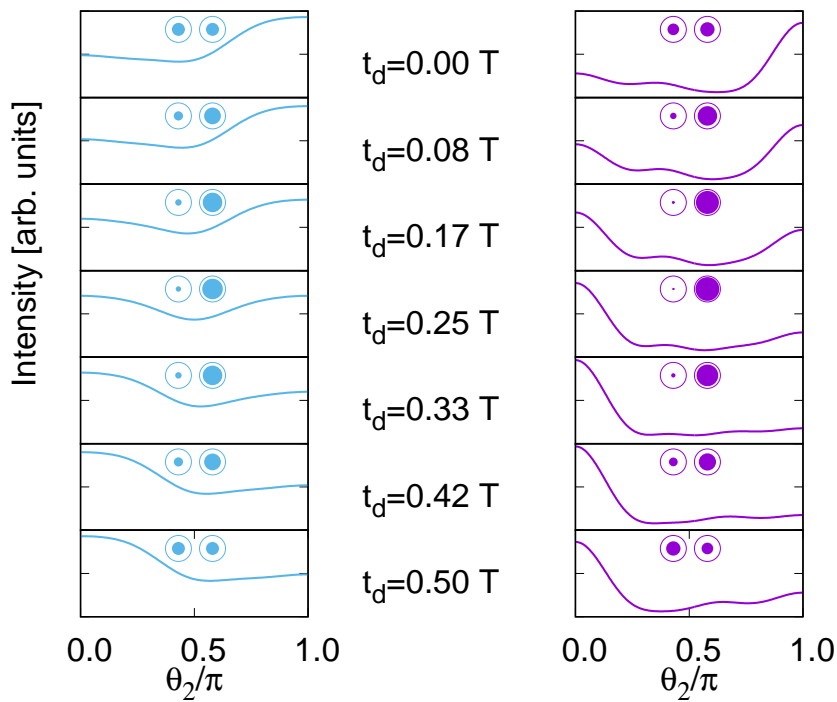}
\caption{Intensity of the second photoelectron for selected delay times $t_d$ as a function of angle $\theta_2$ with respect to the molecular axis for 
the first photoelectron emitted along the molecular axis ($\theta_1=0$)
and considering an average over the orientation of the molecular axis with respect to the polarization axis.
The left panel shows results from the calculation employing the plane-wave approximation,
the right panel shows results from the scattering calculation.
The oscillation of the hole density from the left to the 
right atom is sketched in each sub-plot with the filled circle representing presence of the hole.
\label{fig:angle2}
}
\end{figure}

Our results so far have been based on plane waves for the continuum electrons.
However, at the relatively low photoelectron energies considered ($\unit[15]{eV}$ and $\unit[20]{eV}$), 
the photoelectron wave function can be significantly 
distorted due to the molecular potential.
To test the validity of our results, we have conducted 
scattering calculations for the photoelectron in the potential of the molecular cation or dication that is left behind.
These potentials were calculated in a frozen-core static-exchange approximation~\cite{lucchese_studies_1982,raseev_partial_1980,robb_iterative_1980},
using Hartree-Fock orbitals for \ce{N2} obtained with the Molpro\cite{werner2012molprowires,werner2012molpro} 
quantum chemistry package and an aug-cc-pVTZ basis set\cite{augccpvtz}. 
The photoelectron continuum states have been calculated in the static-exchange Fock potential for the respective orbital configuration making up the core-hole state of interest.
The \texttt{ePolyScat} software package has been used to calculate these 
distributions~\cite{gianturco1994JCP,natalense1999JCP}. 
This package uses the variational Schwinger
method~\cite{lippmann1950PR,blatt1949PR,takatsuka1981PRA,winstead1990PRA,gianturco1994JCP,natalense1999JCP}
on a single-center grid to calculate dipole matrix elements between the
ionic core and the continuum state. 
The resulting angular distributions are plotted in Fig.~\ref{fig:angle2} on the right and compared 
with the plane-wave results on the left panel, which are the same as shown in Fig.~\ref{fig:compare_orient_avg}.
While they show the same qualitative trend as the curves from the plane-wave approximation, some quantitative differences can be seen.
In particular, the degree of orientation of the second photoelectron is much 
stronger pronounced in the scattering calculation as compared to the results based on the plane-wave approximation.
Overall, the emission of the second photoelectron shows in both calculations
a pronounced oscillation following the 
movement of the core hole in the molecule.
The comparison shows that the behavior of the angular correlation 
can be qualitatively understood from the plane-wave approximation that allows one to 
gain an intuitive understanding 
of the connection of core-hole oscillations and the delay-dependent angular correlation of the two photoelectrons. 

As we have demonstrated here,
the variations of the angular correlations illustrated in Figs.~\ref{fig:compare_orient_avg}, \ref{fig:angle12} and \ref{fig:angle2} 
can be directly interpreted as a consequence of the dynamical evolution of the core-hole wave packet given in Eq.~(\ref{eq:corehole_wave_packet}).
Our results show that detecting the molecular-frame angular distribution of the two photoelectrons 
enables one to monitor the oscillation of the core hole shifting from one nitrogen atom to the other.

\subsection{Finite Pulse Overlap \label{sec:finite_overlap}}

\begin{figure}
\subfigure[$t_d=0$]{
  \includegraphics[width=0.49\textwidth]{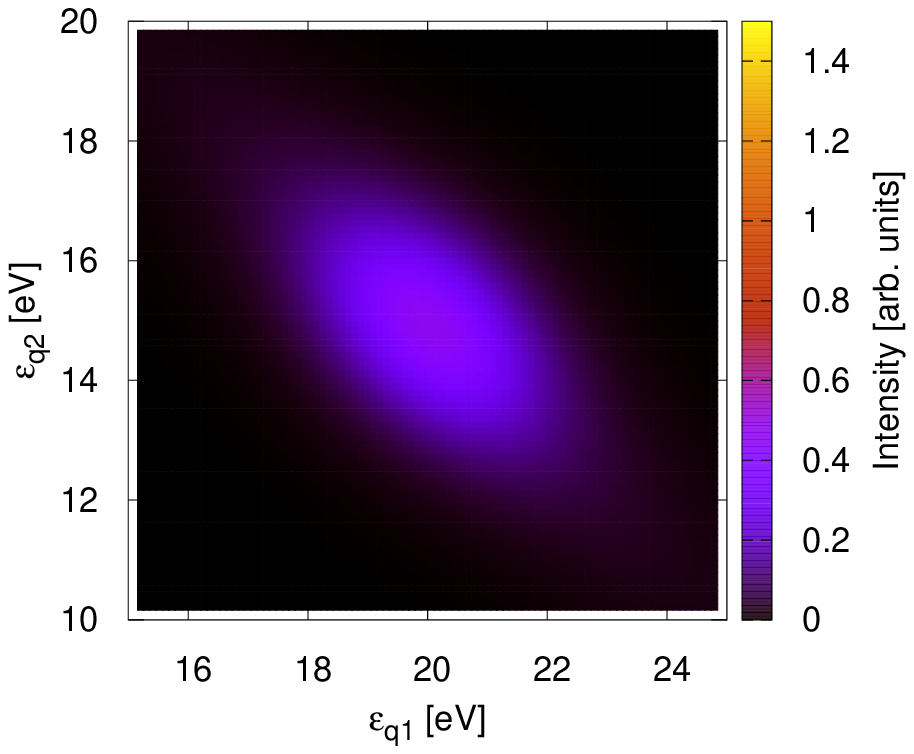}}
\subfigure[$t_d=\sigma$]{
  \includegraphics[width=0.49\textwidth]{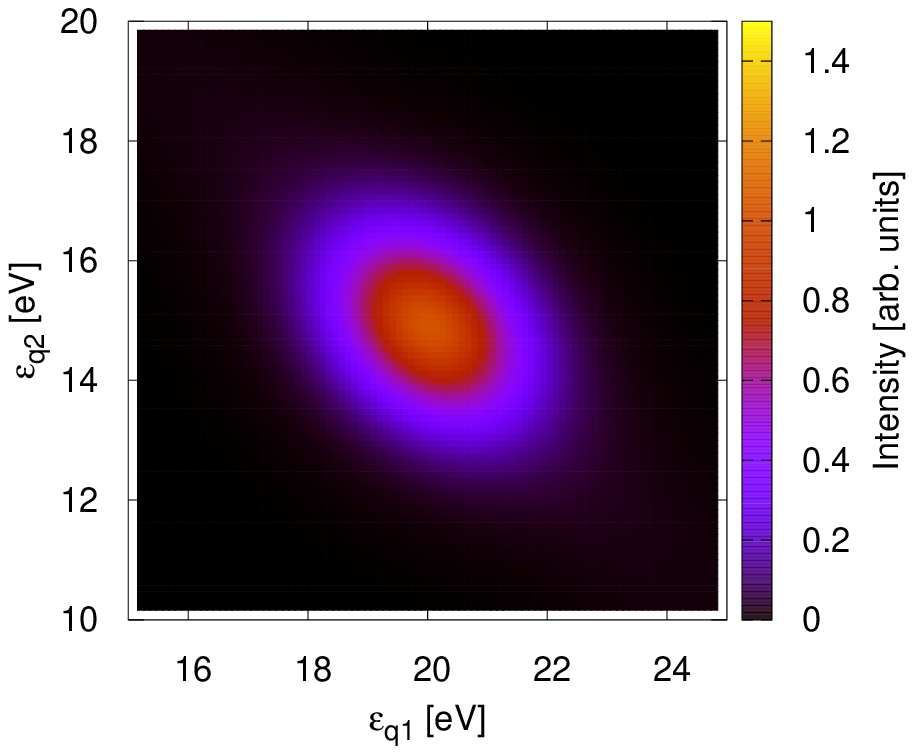}}
\subfigure[$t_d=2\sigma$]{
  \includegraphics[width=0.49\textwidth]{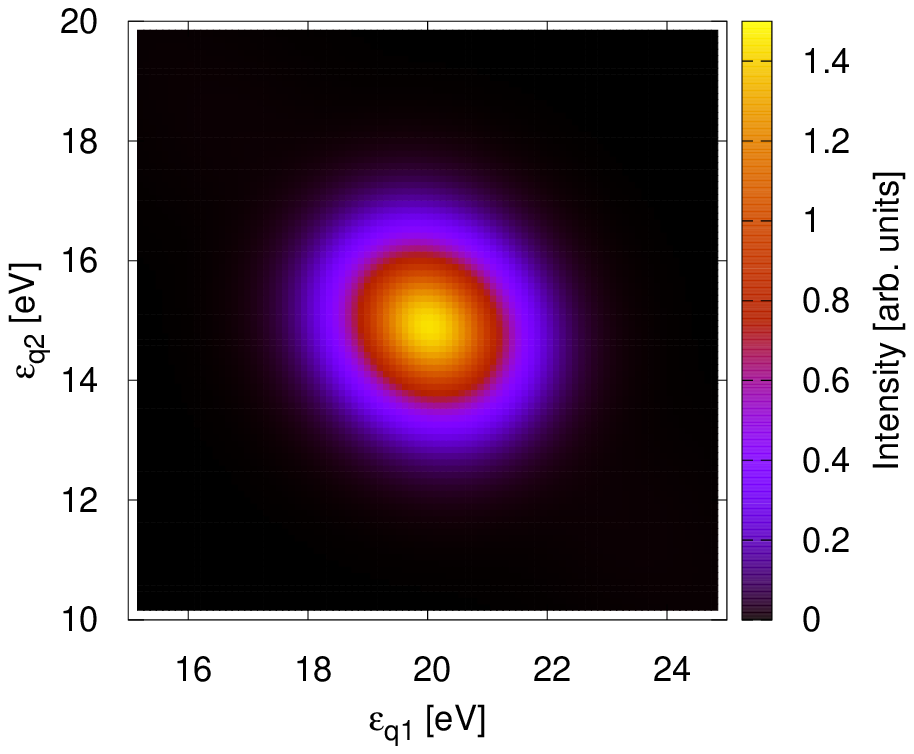}}
\subfigure[$t_d=2\sigma$, assuming no overlap]{
  \includegraphics[width=0.49\textwidth]{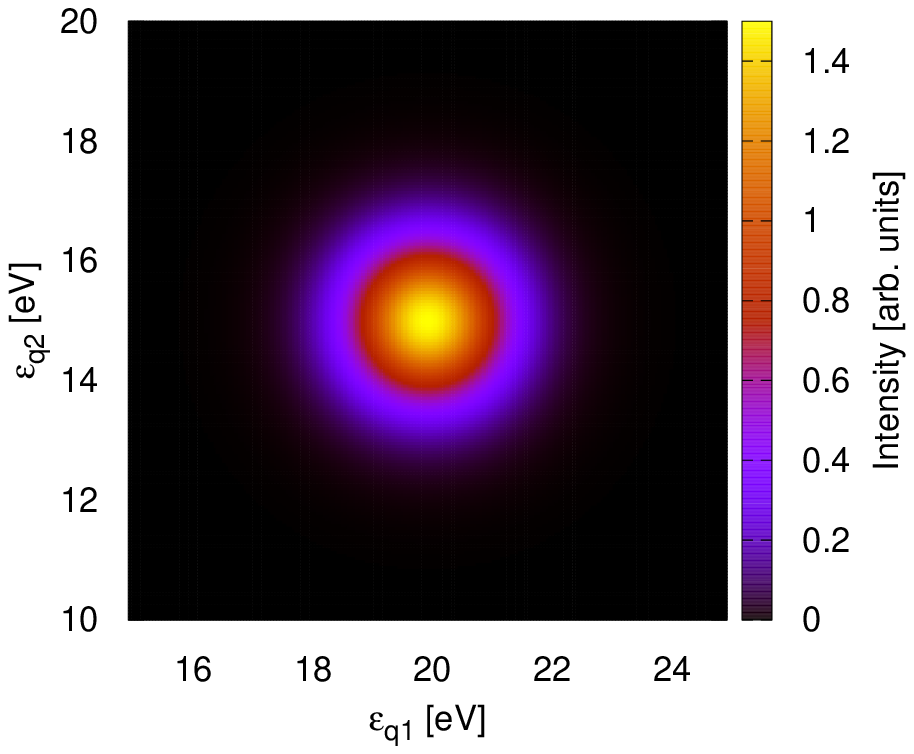}}
\caption{ Two-dimensional spectral line shape of the photoelectrons for emission directions $\theta_1=0$, $\theta_2=\pi$. 
Subfig.~(a-c) take into account effects of finite temporal overlap for different delay times ((a) $t_d=0$, (b) $t_d=\sigma$, (c) $t_d=2\sigma$, where $\sigma$ determines the temporal width of the x-ray pulse).
Subfig.~(d) shows the results for $t_d=2\sigma$, where effects of temporal overal have been neglected.
The same pulse parameters have been used as for Fig.~\ref{fig:compare_orient_avg},~\ref{fig:angle12},~and~\ref{fig:angle2}.
\label{fig:line_shape}
}
\end{figure}

In Eq.~(\ref{eq:timeintegral_simplified}), we have assumed 
negligible temporal overlap of the two pulses.
To inspect the effect for very small delay times $t_d$, 
we discuss here the resulting two-dimensional photoelectron spectrum and 
the angular correlation considering a finite temporal pulse overlap.
To that end, we have numerically evaluated the time integral in Eq.~(\ref{eq:timeintegral}) 
for a pulse duration of $\sigma \simeq \unit[0.5]{fs}$. 
Figure~\ref{fig:line_shape} shows the 
resulting two-dimensional photoelectron spectral line shape for different delay times  
$t_d=0,\sigma,2\sigma$ evaluated for the angle pair $\theta_1=0, \theta_2=\pi$ and 
the same pulse parameters as used in Figs.~\ref{fig:compare_orient_avg},~\ref{fig:angle12},~and~\ref{fig:angle2}.
For comparison, the last sub-figure shows the two-dimensional line-shape at $t=2 \sigma$ evaluated using the 
no-overlap approximation (Eq.~(\ref{eq:timeintegral_simplified})), which is a pure two-dimensional Gaussian with spectral
width $1/\sigma$ in $\epsilon_{q_1}$ and $\epsilon_{q_2}$.

As can be seen, the effect of temporal overlap
leads to a significant reduction of the line intensity. 
This reduction results from the lower probability of 
having the photoabsorbtion in the expected order (first absorption by the first pulse, second absorption in the second pulse).
Furthermore, the spectral line shape is distorted
towards an anti-correlated shift of the two photoelectron energies.
This can be rationalized 
by conducting the integral in $I_{12}(\Delta E_1,\Delta E_2)$ in Eq.~\ref{eq:ppn3} for $t_d=0$, which yields (assuming $\Gamma \sigma \simeq 0$)
\begin{equation}
I_{12}(\Delta E_1,\Delta E_2)=
I(\Delta E_1) I(\Delta E_2) 
\text{erfc} \left( \frac{i \sigma }{2} (\Delta E_1 -\Delta E_2) \right),
\end{equation}
where $\text{erfc}(x)$ is the complementary error function.
The $\text{erfc}( i \sigma /2 (\Delta E_1 -\Delta E_2) )$ factor increases along $|\Delta E_1 -\Delta E_2|$
and thus gives rise to a broader distribution along $|\Delta E_1 - \Delta E_2$|.

\begin{figure}
\subfigure[$t_d=0$]{
  \includegraphics[width=0.4\textwidth]{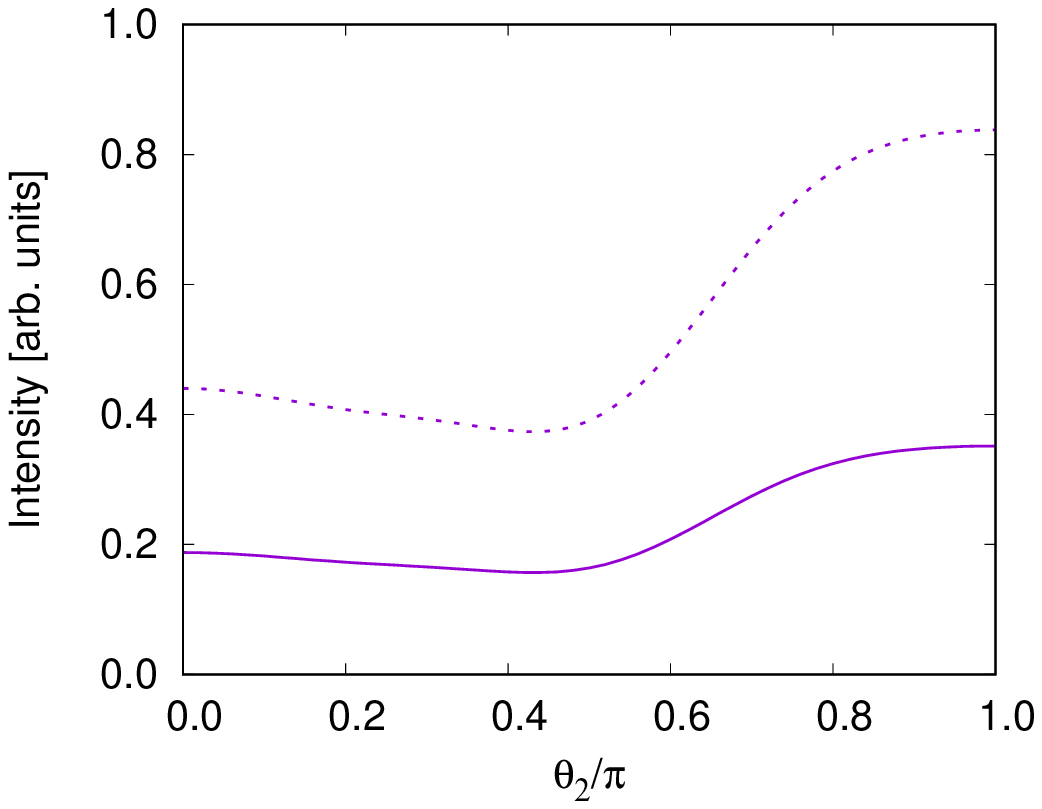}}
\subfigure[$t_d=\sigma $]{
  \includegraphics[width=0.4\textwidth]{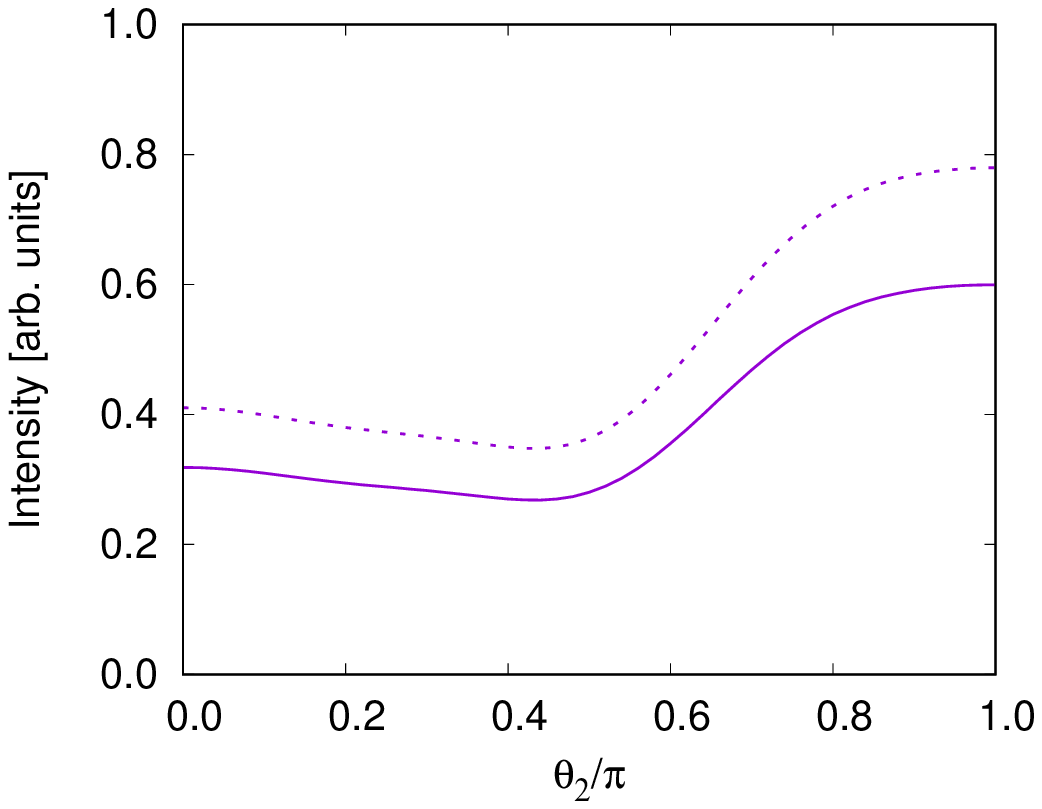}}
\subfigure[$t_d=2\sigma$]{
  \includegraphics[width=0.4\textwidth]{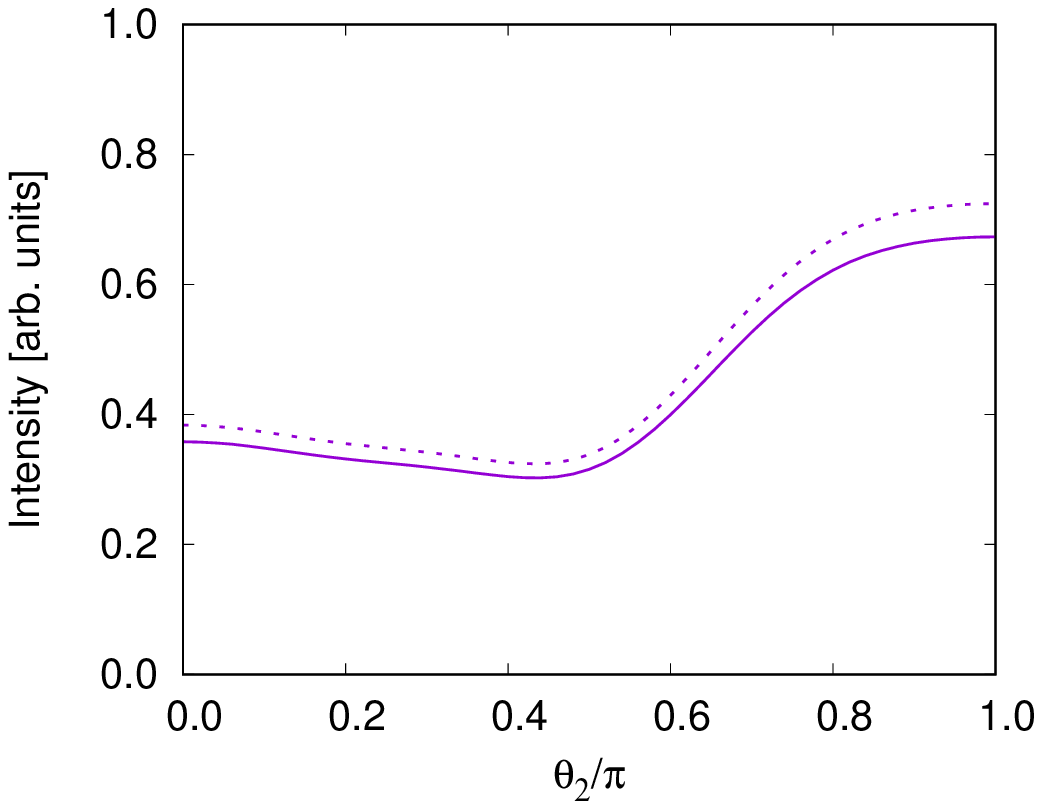}}
\subfigure[$t_d=3\sigma$]{
  \includegraphics[width=0.4\textwidth]{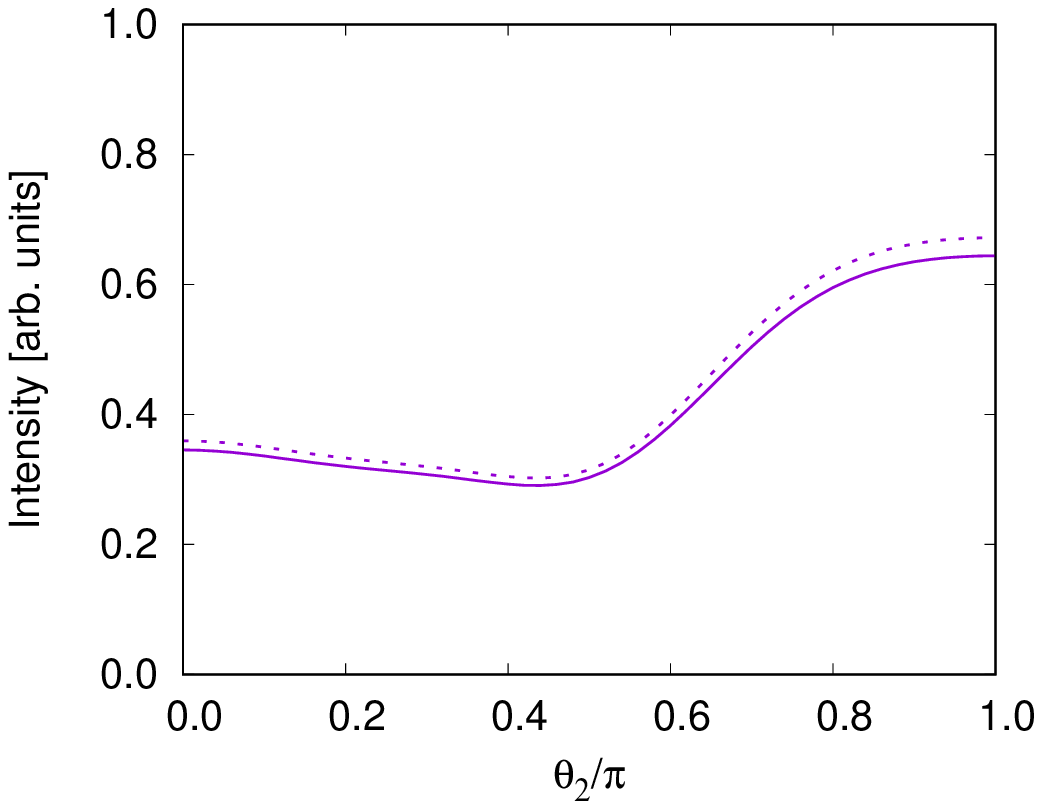}}
\caption{
  Intensity of the second photoelectron for selected delay times $t_d$
  ( (a)~$t_d=0$, (b)~$t_d=\sigma$, (c)~$t_d=2\sigma$, and (d)~$t_d=3\sigma$)
  as a function of angle $\theta_2$ with respect to the molecular axis for the first photoelectron emitted along the molecular axis ($\theta_1=0$)
and considering an average over the orientation of the molecular axis with respect to the polarization axis.
The solid line shows results taking into account the finite temporal overlap of the pulses.
The dotted line shows the result assuming no temporal overlap.
Results are based on the plane-wave approximation for the two photoelectrons.
\label{fig:angle2_overlap}
}
\end{figure}

To investigate the effect of finite overlap on the angular distribution of the two photoelectrons, 
Fig.~\ref{fig:angle2_overlap} shows the energy-integrated intensity as a function of $\theta_2$ for $\theta_1=0$ and selected delay times.
The curves indicate a reduced intensity at finite temporal overlap 
of the pulses as observed before in Fig.~\ref{fig:line_shape}. 
Apart from this effect, no qualitative difference is seen for the angular dependence.
Our analysis indicates that the finite temporal overlap of the pulses has no qualitative effect 
on the angular correlation patterns discussed here. 
Apart from some distortion of the spectral line shape, 
the temporal overlap of the pulses simply results in an overall reduction of the two-photoelectron signal.

\section{Feasibility \label{sec:realization}}
Several experimental challenges have to be overcome to measure the predicted signatures of coherent core-hole dynamics. 
First, the experiment requires two sub-femtosecond x-ray pulses with two different photon energies above the nitrogen K edge, 
sufficient (coherent) bandwidth to cover the 100-meV splitting between the 1$\sigma_g$ and 1$\sigma_u$ levels, 
sub-femtosecond adjustable delay between the pulses, and enough pulse energy/fluence to (nearly) saturate the single-photon ionization of the N$_2$ core level, i.e., photon fluence exceeds 1 photon per cross section.
Second, the two photoelectrons produced by each of the x-ray pulses have to be measured in coincidence, and the angle between the two photoelectrons should be measured in the molecular frame, i.e., with respect to the direction of the molecular axis (See Sec.~\ref{sec:theory}). 
And third, the photoelectrons stemming from the ionization of an excited N$_2^+$ molecule with a single core-hole by the 
second x-ray pulse have to be 
distinguished from those produced by competing single and two-photon processes, 
some of which are rather close in kinetic energy and may be produced at a higher rate.
While these requirements are not (yet) met by existing x-ray sources, the proposed experiment should be feasible in a year or two, 
once the upgraded LCLS-II high-repetition rate x-ray free-electron laser at SLAC and the associated electron-ion coincidence 
momentum imaging end-station in the Time-resolved Atomic, Molecular and Optical Science (TMO) hutch are operational, both of which are currently under construction. 
The possibility to produce intense, two-color, few-femtosecond x-ray pulses with adjustable delay was already demonstrated at 
LCLS-I\cite{lutman_experimental_2013,picon_heterositespecific_2016}, as was the generation of sub-femtosecond x-ray pulses\cite{huang_generating_2017}. 
In particular, in the soft x-ray domain, recent experiments have demonstrated the production of single attosecond x-ray pulses at 600 eV photon energy with pulse durations as short as 480 as, a coherent bandwidth of 5 eV (FWHM), and pulse energies up to 50 $\mu$J\cite{duris_tunable_2019}.
Attosecond soft x-ray pulses in combination with the two-color two-pulse capability will also be available at LCLS-II.

Given the partial K-shell photoionization cross section of nitrogen molecules of approximately 1 Mbarn at 20 to 30 eV above the ionization threshold, a pulse energy (on target) of $\unit[430]{eV} \times \unitfrac[1]{\mu m^2}{Mb} < \unit[1]{\mu J}$ is necessary to saturate the single-photon ionization process if the x rays are focused to a spot size of 1 micron by 1 micron. 
Assuming a total ionization rate of 0.3 events per x-ray pulse in order to maintain clean coincidence conditions, a (sequential) two-color ionization event will occur in 9 out of 100 x-ray shots (consisting of two attosecond pulses per shot). 
For the case of N$_2$ inner-shell ionization, where the axial-recoil approximation is known to be valid to a good approximation\cite{weber_kshell_2001}, coincident and momentum-resolved detection of two photoelectrons and two fragment ions will enable the measurement of the emission direction of both photoelectrons in the molecular frame. 
For photoelectron kinetic energies of 20 to 30 eV and $N^{2+}$ fragment ion kinetic energies of up to 40 eV, detection of both photoelectrons and both fragment ions in the full solid angle will be possible with the planned electron-ion coincidence spectrometer at LCLS\footnote{Peter Walter, private communication (2019)}. 
Assuming an electron and ion detection efficiency of 40\% (limited mostly by the efficiency of the microchannel plate detectors 
and the transmission of meshes in the spectrometer), the total detection efficiency for four-particle coincidences will be $0.4^4 = 2.6\%$. 
At an x-ray pulse repetition rate of 100 kHz, approximately $100,000 \times 0.3^2 \times 0.4^4= 230$ four-particle coincidence events per second stemming from sequential 
two-color ionization will thus be detected. In order to measure the yield of coincident electrons as a function of their emission angles, 
the detected events have to be binned into approximately 30 angular bins for each electron 
(corresponding to cones with a 6 degree opening angle and taking advantage of the cylindrical symmetry if we average over all orientations of the molecular axis 
with respect to the x-ray polarization axis). Assuming (nearly) isotropic electron emission, 
this will require approximately one hour of data acquisition for each delay point to reach 1000 coincidence events in a given electron-electron angular bin.
At small delays, the vast majority of these events will correspond to emission of the second K-shell photoelectron before the core hole has decayed. 
At larger delays, this fraction is reduced exponentially due to the competing Auger decay, such that after a quarter period (approx. 10 fs), only 36\% of the second K-shell photoelectron emission events occur before the Auger decay.

Even under these favorable count-rate conditions, a remaining challenge will be to spectrally separate the two-photon signal of interest, i.e., the photoelectrons stemming 
from the ionization of an excited N$_2^+$ molecule with a single core hole by the second x-ray pulse, from other competing one-~and two-photon processes. 
This requires a careful choice of photon energies in order to avoid spectrally overlapping contributions. 
For the photon energies assumed above (430 eV for the first pulse and 440 eV for the second pulse), 
the photoelectron produced by the first pulse will have a kinetic energy of 20 eV, 
and the photoelectron produced by the ionization of an excited N$_2^+$ molecule with a single core-hole by the second x-ray pulse will have a kinetic energy of 15 eV.
Given a photon bandwidth of approximately 5 eV (FWHM), both photoelectron peaks might slightly overlap in the experimental photoelectron spectrum, but the features should be clearly resolvable
if the photon energy of the first pulse is increased slightly to $\unit[433]{eV}$. In that case the photoelectrons produced by the second pulse can still
be distinguished from the second-step 
photoelectrons produced by sequential two-photon ionization within the first pulse, which have a kinetic energy of 8 eV, if a photon energy of $\unit[433]{eV}$ is chosen. 
Note that some of the core-ionized molecules produced by the first pulse will have already decayed via Auger decay when the second photon is absorbed, in which case the second ionization will occur in an 
(N$_2$)$^{2+}$ molecular dication or, for longer pump-probe delays, in an N$^+$ fragment ion, 
which is formed after Auger decay and subsequent fragmentation of the molecule. Those two ions have N(1s) ionization energies of approximately 440 and 432 eV, 
respectively, and the resulting photoelectrons will thus have kinetic energies of approximately 0 and 8 eV, respectively. 
All of the expected spectral contributions 
should thus be well resolvable in the proposed experiment, and further separation of competing contributions and suppression of unwanted signal is possible through additional gating on the four-particle coincidence spectra, even in case the features in the noncoincident electron spectra are slightly overlapping.

\section{Conclusion \label{sec:conclusion}}
We have demonstrated that the electronic wave-packet dynamics
initiated by core-ionizing \ce{N2} with a large bandwidth sub-femtosecond x-ray pulse
can be monitored via angle-resolved double-core-hole spectroscopy.
By detecting the two photoelectrons created by the two sub-fs x-ray pulses 
in coincidence, we can monitor the oscillations of the core hole as it goes from one nitrogen to the other.
By assuming a simple plane-wave model for the continuum electrons,
analytical relations have been derived that facilitate an intuitive understanding of the
relation between the core-hole dynamics and the resulting dynamics in the photoelectron angular distribution.
For the case of photoionization into a TSDCH state, 
we have discussed the resulting angular distribution and validated our results with
a more elaborate scattering calculation for the photoelectron wave function.
Furthermore, we have discussed the feasibility of the experiment in light of the upcoming 
technical improvements at XFEL facilities.

Being able to monitor the core-hole wave-packet dynamics in molecules 
will give us unprecedented insight into the electronic dynamics following
core ionization and will allow us to validate our theoretical
understanding of the attosecond response following
x-ray ionization of molecules.
In the current work we focus on the core-hole wave-packet dynamics in \ce{N2}.
We note, however, that the proposed scheme
is also applicable to studying core-hole wave-packet dynamics in other molecules, e.g.,
coherent superpositions of core holes in hydrocarbon chains.
The central challenge here is to have pulses with a coherent bandwidth large 
enough to cover the core-ionized eigenstates of interest.
A particularly interesting generalization could be elucidating the core-hole wave-packet dynamics 
in the molecules \ce{C_2H_2}, \ce{C_4H_4}, and \ce{CO_2}.
Core ionization in these molecules triggers interesting dynamics,
because vibrational coupling through nonadiabatic effects gives rise to dynamical 
core-hole localization\cite{domcke_vibronic_1977,kivimaki_vibrationally_1997,rescigno_tracking_2015}.
Another generalization is to study the valence electron dynamics that accompanies core ionization through shake-up excitations\cite{kuleff_core_2016}. 
The energetic separation of the involved electronic eigenstates is, 
however, much larger (several eV) than in the scenario considered here, 
and therefore poses much higher challenges on the x-ray pulse characteristics.
Our work addressing the coherent core-hole wave-packet dynamics in \ce{N2} 
can be seen as a pioneering step in these directions.
By proposing this kind of experimental scheme
and discussing its feasibility in detail,
the findings presented here are thus relevant
for the design and interpretation of a wide range of prospective experiments.

\section{Acknowledgements}
The authors thank Siddhartha Chattopadhyay for carefully checking the equations.
This work was supported by the Chemical Sciences, Geosciences, and Biosciences Division, Office of Basic Energy Sciences, Office of Science, US Department of Energy, Grant No. DE-SC0019451.
This research used resources of the National Energy Research Scientific Computing Center (NERSC), a U.S. Department of Energy Office of Science User Facility operated under Contract No. DE-AC02-05CH11231.

%


\end{document}